\documentclass{sig-alternate}
 \pdfoutput=1

\makeatletter
\let\note\@undefined        
\makeatother

\usepackage{graphicx,color}
\usepackage{amsmath}
\usepackage{amsfonts}
\usepackage{amssymb}
\usepackage{amsthm}
\usepackage{url} 
\usepackage{multicol}
\usepackage{multirow}
\usepackage{setspace} 
\usepackage{xspace}
\usepackage{listings}
\usepackage{ifthen}
\usepackage{verbatim}
\usepackage{float} 
\usepackage{mathptmx}
\usepackage{stmaryrd} 
\usepackage{textcomp}
\usepackage{todos} 
\usepackage{fixltx2e} 
\usepackage{booktabs}
\usepackage{fancyhdr}

\usepackage[mathletters]{ucs}
\usepackage[utf8x]{inputenc}
\usepackage{wasysym}	

\usepackage[bookmarks, colorlinks, citecolor=green, urlcolor=blue,
			filecolor=blue, linkcolor=blue]{hyperref}
\usepackage[noend]{algorithmic} 
\usepackage{algorithm}

\newcommand{\ie}{\hbox{\emph{i.e.}}\xspace}

\newfloat{Protocol}{thp}{lop}
\newfloat{Program}{thp}{lop}
\newfloat{Procedure}{thp}{lop}

\theoremstyle{plain}
\newtheorem{lemma}{Lemma}
\newtheorem{theorem}{Theorem}[section]

\theoremstyle{definition}
\newtheorem{definition}{Definition}[section]

\DeclareUnicodeCharacter{183}{\cdot}						
\DeclareUnicodeCharacter{931}{\ensuremath\Sigma}			
\DeclareUnicodeCharacter{9001}{\ensuremath\langle}			
\DeclareUnicodeCharacter{9002}{\ensuremath\rangle}			
\DeclareUnicodeCharacter{9608}{\ensuremath\blacksquare}		
\DeclareUnicodeCharacter{1013}{\in}							
\DeclareUnicodeCharacter{8213}{---}							

\usepackage[normalem]{ulem}	

\usepackage{paralist}
\usepackage{caption}
\DeclareCaptionType{copyrightbox}
\usepackage[T1]{fontenc}
\usepackage{framed} 
\usepackage{comment}
\usepackage{wrapfig}
\usepackage{fixltx2e} 
\usepackage{lipsum}
\usepackage{mathpartir}
 \usepackage{amssymb}

\usepackage{graphicx}
\usepackage{caption}
\usepackage{subcaption}

\usepackage{mathptmx}  
\usepackage{amsmath}
\usepackage{amsthm}
\usepackage{pifont}

\newlength{\emstr}
\newcommand{\thetitle}{
  Detecting Malware with Information Complexity
}

\newcounter{RQCounter}

\hypersetup{
  pdftitle={\thetitle},
	pdfauthor={Nadia Alshahwan, Earl T. Barr, David Clark, George Danezis},
  plainpages=false
}

\title{\thetitle}

\author{%
  \qquad
Nadia Alshahwan 
\quad Earl T. Barr 
\quad David Clark
\quad George Danezis
\qquad
\and 
\begin{tabular}{c}
  \\
  \affaddr{Department of Computer Science} \\
  \affaddr{University College London,  UK} \\
  \texttt{nadia.alshahwan,e.barr,david.clark,g.danezis@ucl.ac.uk}
\end{tabular}
}

\begin{document}

\maketitle

\begin{abstract}
This work focuses on a specific front of the malware detection arms-race, namely  the detection of persistent, disk-resident malware.  We exploit normalised compression distance (NCD), an information theoretic measure, applied directly to binaries.  Given a zoo of labelled malware and benign-ware, we ask whether a suspect program is more similar to our malware or to our benign-ware. Our approach classifies malware with 97.1\% accuracy and a false positive rate of 3\%. We achieve our results with off-the-shelf compressors and a standard machine learning classifier and without any specialised knowledge. An end-user need only collect a zoo of malware and benign-ware and then can immediately apply our techniques.

We apply statistical rigour to our experiments and our selection of data. We demonstrate that accuracy can be optimised by combining NCD with the compressibility rates of the executables. We demonstrate that malware reported within a more narrow time frame of a few days is more homogenous than malware reported over a longer one of two years but that our method still classifies the latter with 95.2\% accuracy and a 5\% false positive rate. Due to the use of compression, the time and computation cost of our method is non-trivial. We show that simple approximation techniques can improve the time complexity of our approach by up to 63\%. 

We compare our results to the results of applying the 59 anti-malware programs used on the VirusTotal web site to our malware. Our approach does better than any single one of them  as well as the 59 used collectively. 
 
\end{abstract}

\section{Introduction}
\label{sec:intro}

Arms races are often ruinous.  The malware arms race is no exception.  Despite the widespread use of anti-virus software, malware is imposing a significant productivity tax on society, slowing machines and wasting bandwidth. Moreover, there are no SALT talks to provide respite periods. The race is relentless.

The evolving sociology of malware, and in particular the growth of industrial scale production of polymorphic and metamorphic variants of existing malware, is straining the ability of existing methods of detection, via bit signatures and dynamic analysis, to cope with this production volume. A number of researchers have considered ways to leverage existing techniques. Since the early to mid 2000's there has been research into semantics of programs, on the assumption that identifying semantic invariants can help to overcome metamorphic variation.  

This decade has seen research into similarity metrics and applications of machine learning. The motivation is that signature-based approaches, whether syntactic or behavioural, require painstaking manual
analysis to isolate the signature, and so cannot handle previously unseen
malware. In contrast, a successful similarity metric could extrapolate the features of its two labeled
input sets of malware and non-malware and, at least some of the time, avoid upfront analysis and
detect previously unknown malware, potentially automatically.

A generic similarity metric has existed for the last decade or so, the Normalised Information Distance (NID)~\cite{LCL+SM2004}. It works on syntax rather than semantics but it has a deep mathematical foundation that is connected to information theory, probability theory, the theory of randomness, and algorithmic complexity theory. It has the unparalleled advantage of being universal in the sense that it minorises (intuitively: incorporates) every other possible similarity measure --- but has the slight drawback of being uncomputable. On the other hand there exists an effective approximation to NID that uses compressors, the Normalised Compression Distance (NCD)~\cite{CilibrasiV05}.

The promise of building a malware detector on the back of NCD is its good approximation to universality.  If NCD detects malware well, it could do so without requiring any static or dynamic analysis or preprocessing.  All one would need to detect malware is sufficient processing power, a good compressor, and known collections of malware and non-malware, and nothing more. Being generic, it can be directly applied to binary executables. This lack of need for specialised knowledge and techniques introduces a ``garage assembly'' element into the approach. 

Our objective in this paper has been to answer the question, ``How well does NCD detect malware when we apply it to binary executables?", and to answer it with empirical rigor. The answer, incidentally, is, ``Very well!''. We found that when you apply the method to malware collected within a short time scale of a few days it detects malware with 97.5\% accuracy --- rather astonishing from a standing start. Once you attain accuracy levels in the mid 90's, it becomes increasingly difficult to improve on them so we used the flexibility and power of the Decision Forest Classifier to improve accuracy incrementally to this 97.5\% level.

In this paper we have made the following contributions:

\begin{enumerate}

\item We have conducted the first statistically rigorous, experimental evaluation of the ability of NCD to detect malware using only binary executables.

\item  We have demonstrated that NCD, as used in our approach, is competitive with commercial anti-malware tools:  it outperforms any single one and matches the performance of all of them together.

\item We have used NCD in conjunction with a state-of-the-art classifier, the Decision Forest Classifier, that scales well, is inherently parallel, and is tuneable.

\end{enumerate}

We have been inspired by Wehner's 2007 paper in which she applies NCD to cluster polymorphic worms found in network traffic \cite{Weh+JCS2007}. Unlike our own work, her work lacked any attempt to make a definitive statistical statement.

\section{Background}
\label{sec:background}

This paper is founded on the use of compressors to achieve upper bounds on the Kolmogorov complexity of strings. (This concept of complexity is also called 
algorithmic information  or information complexity.) Our interest is in the application of an existing universal, generic, similarity metric to detect malware. This similarity metric is called the Normalised Information Distance (NID). Like Kolmogorov complexity it is not computable so we are forced to use the best, general approximation to it, called the Normalised Compression Distance (NCD).


 In what follows we outline the underlying ideas, try to provide intuitions about how the similarity metric works, and outline the relationship between Kolmogorov complexity and NCD. The material paraphrases material in papers by Li, Cilibrasi, Vit\'anyi  and others and further readings may be
found in Li and Vitanyi's book \cite{LCL+SM2004,LV+IKCA2008}.

First, we work in a world of strings so any objects we wish to consider must be encoded as strings. This is not a strong restriction as numbers, computer programs and many other objects can be encoded as strings. So, encode all objects as binary strings $x \in \{0,1\}^*$. We can totally
order all such strings, first according to length, and then lexicographically within
each length. Every string in the ordering, x, can then be identified with the
number of its position in the ordering;  its length can be given by a function 
len where
\begin{align*} 
\mbox{len}(x) = \lfloor \mbox{log}(x+1)\rfloor.
\end{align*} 
It is desirable that we can work, if possible, in a setting where the set of strings we consider is a prefix set, i.e. no string in the set is the prefix of another. This restriction is not necessary but desirable. The prefix set property tends to imply other good properties, for example being able to
associate a probability distribution (technically a semi-measure) with our binary strings -- which in turn assists in making the connection between Kolmogorov Complexity and
Shannon entropy. One way to obtain a prefix-free encoding of numbers and programs is to use a self-delimiting code for a string, $x$, such as 
\begin{align*}
  \overline{x} = 1^{\mbox{len}(x)}0x
\end{align*} 
Then it can be shown that $\{\overline{x}: x \in \{0,1\}^*\}$ is prefix-free~\cite{LV+IKCA2008}. 

Kolmogorov Complexity is sometimes called the universal minimum description
length. The act of description requires a description language so we say that
$\varphi(p)=x$ means that $p$ is a description of $x$ using the description
language $\varphi$. Now we can define the conditional complexity of a string
given a starting string as the length of the shortest description that takes
the given, starting string as input and produces the string.

\begin{definition}(The conditional complexity of a string)

The conditional complexity of $x$ given $y$ is \[C_{\varphi}(x|y) =
\mbox{min}\{\mbox{len}(p): \varphi(y,p) = x\}\]

\end{definition}

Here, $p$ is a partial recursive function that takes $y$ as input and outputs $x$. The conditional complexity  of $x$ is just the length of the shortest funtion that can do this

It can be shown that the set of partial recursive functions on a domain of
prefix-free strings is sufficiently expressive to capture description languages
and that for every pair of strings there is, in theory, a best description language in this set in that
descriptions in this language are shorter than descriptions in any other
language for this pair of strings.

\begin{theorem}

$\exists$ a partial recursive prefix function $\Psi_0$ s.t. $\forall$ partial
recursive prefix functions $\Psi$ there is a constant $c$ with
\[C_{\Psi_0}(x|y) \le C_{\Psi}(x|y) + c\]
for all pairs of strings $x$ and $y$. The constant $c$ is independent of $x$ and $y$, only depending on $\Psi_0$ and $\Psi$. 
\end{theorem}

The import of this theorem is that minimum description length is language dependent -- but by fixing the description language we do not lose out.  As we consider increasingly longer strings, the concept of a minimal description length independent of the description language asserts itself. One consequence is that when we apply our similarity metric to malware, confidence in our results increases with the length of the strings compared.

The conditional Kolmogorov complexity of a string $x$ given a string $y$ is written
$K(x|y) = C_{\Psi_0}(x|y)$ while the (unconditional) Kolmogorov Complexity of a
string is written $K(x) = C_{\Psi_0}(x|\epsilon)$ and $\Psi_0$ is usually taken to be a universal computer such as the universal Turing machine.

The similarity metric ideally should be a distance metric, i.e. should satisfy the axioms of a distance metric. The obvious definition for the minimal information distance between a pair of
strings, $x$ and $y$, is the length of the shortest program for a universal
computer to transform $x$ into $y$ and $y$ into $x$. However the conditional
algorithmic complexities of two strings considered in each order are not in
general equal, i.e. $K(x|y) \not = K(y|x)$. Because symmetry fails, this
``obvious'' definition is not a distance metric. This can be fixed by simply
taking the least upper bound of the conditional complexities. Then there remains
the problem of comparing strings of quite different lengths. 
``Normalising'' the distance with respect to the least upper bound of
the algorithmic complexities of the two strings handles this problem. The result 
is the Normalised Information Distance (NID). 

\begin{definition}(NID)
\begin{align*}
e(x,y) = \frac{\mbox{max}\{K(x|y),K(y|x)\}}{\mbox{max}\{K(x),K(y)\}}
\end{align*}

\end{definition}

NID calculates a value in $[0,1]$ and is a universal, generic, upper semi-computable distance metric satisfying a
density requirement in $\{0,1\}^*$. It is universal because it can be shown to
be less than any other similarity metric between two strings. So if any two
strings are similar because of any feature that they share and this can be
captured in a metric, it can also be captured with NID. This universality in
turn makes NID completely generic. It does not depend on any particular features of the strings so it can be applied to any type of strings. It is upper semi-computable because it can
be approximated from above by a sequence of functions into the rational numbers
that converge on NID in the limit. It is a useful, non-degenerate metric because at any finite distance from a string there is at most a certain, finite number of other strings. 

However, Kolmogorov complexity is not a partial recursive function so it is not computable and must be approximated. Any approximation is necessarily an upper bound, as explained above. We use compression programs to calculate computable upper bounds. Compressors are not the only possible way to approximate algorithmic information content but they are natural, since they exploit repetitive patterns in a string. The better your compression program, the more tightly your upper bound approximates Kolmogorov complexity. Useful detection and classification methods for malware using algorithmic complexity give an advantage in the arms race as to do better may require the intellectually prohibitive cost of developing a better compressor. Even then, it is not immediately clear how this advantage can be exploited.

We follow Cilibrasi and Vitanyi in simply replacing $K(x)$ and  with $Z(x)$ where $Z(x)$ is the length of the compressed version of string $x$ produced by a compression program $Z$. Simply substituting $Z$ for $K$ in the definition of NID creates the problem of interpreting $Z(x|y)$. To sidestep this problem, Cilibrasi and Vitanyi employ the following result.

\begin{lemma}
\[\mbox{max}\{K(x|y),K(y|x)\} = K(xy) - min\{K(x), K(y)\}\]
where $xy$ is the concatenation of the strings $x$ and $y$.

\end{lemma}

The result of replacing $K$ with the upper bound $Z$ is called the Normalised Compression Distance

\begin{definition}(NCD)
The Normalised Compression Distance is given by
\[e_Z(x,y) = \inferrule
	                    {Z(xy) - \mbox{min}\{Z(x),Z(y)\}}
	                    {\mbox{max}\{Z(x),Z(y)\}} \]
\end{definition}

Like NID, NCD calculates a value in $[0,1]$ although this outcome depends on how \emph{normal} is the compressor used. The key term in the definition is the term $Z(xy)$ as this is what makes the distance work. It is also the term that is more expensive to compute as it's a longer string to compress than $x$ and $y$ individually and because the number of comparisons grows quadratically with respect to the number of strings being compared.  A \emph{normal} compressor is so called because it is well behaved with respect to this term.

\begin{definition}{Normal Compressor}
A compressor, $Z$, is \emph{normal} if it satisfies for all strings $x$, $y$ and $z$:
\begin{enumerate}
\item $Z(xx) = Z(x)$ and $Z(\epsilon)=0$,
\item $Z(xy) \ge Z(x)$,
\item $Z(xy) = Z(yx)$,
\item $Z(xy) + Z(z) \le Z(xz) + Z(yz)$
\end{enumerate}
up to an additive $O(\mbox{log } n)$ term with $n$ the maximal binary length of any string involved in the (in)equality. 
\end{definition}

Assuming that we use a normal compressor, intuitions about how the distance
works can be gained by applying NCD to a single file. $Z(xx)=Z(x)$ so NCD$(x,x)
= Z(x)-Z(x)/Z(x)=0$, i.e. every file is completely similar to itself.  Not all
compressors behave in this \emph{normal} way. We discuss our choice of 7zip as
compressor in \autoref{sec:approach}.  

To build your intuition about how NCD works, it is instructive to consider
$NCD_S$, which replaces
the normal compressor $Z$ with file size in the definition of NCD.
$\text{NCD}_S$ has the unwanted property that $\text{NCD}_S(x,y) = 1$, for
all $x$ and $y$.  WLOG, let $S(x) < S(y)$, then 
\begin{align*}
NCD_S  = & \frac{S(xy) - \min(S(x),S(y))}{\max(S(x),S(y))} 
= \frac{S(x) + S(y) - S(x)}{S(y)} = \frac{S(y)}{S(y)} = 1.
\end{align*}

\section{Classifying Malware Using NCD} 
\label{sec:approach}

The Normalised Compression Distance (NCD) works by finding common patterns
between two strings using compression algorithms. Therefore, there are no
restrictions on the type of strings to which we can successfully apply NCD. If we want to use NCD to find similarities between programs, we can apply it to any string representation of a program, such as the source code, an execution trace, an abstraction of the program or the program's binary. 

Applying NCD directly to the binary representation of the program eliminates the
manual effort needed to reverse engineer or execute the program to obtain an
execution trace. In the case of malware, this is particularly useful because
malware writers go to great lengths to prevent their programs from being
reversed engineered or revealing their malicious behaviour when executed within
a controlled environment like a virtual machine. Further, executing malware outside of a controlled environment is unsafe.  

\subsection{Choice of Compressor}

\begin{figure}[t]
  \begin{center}
    \includegraphics[width=0.95\columnwidth]{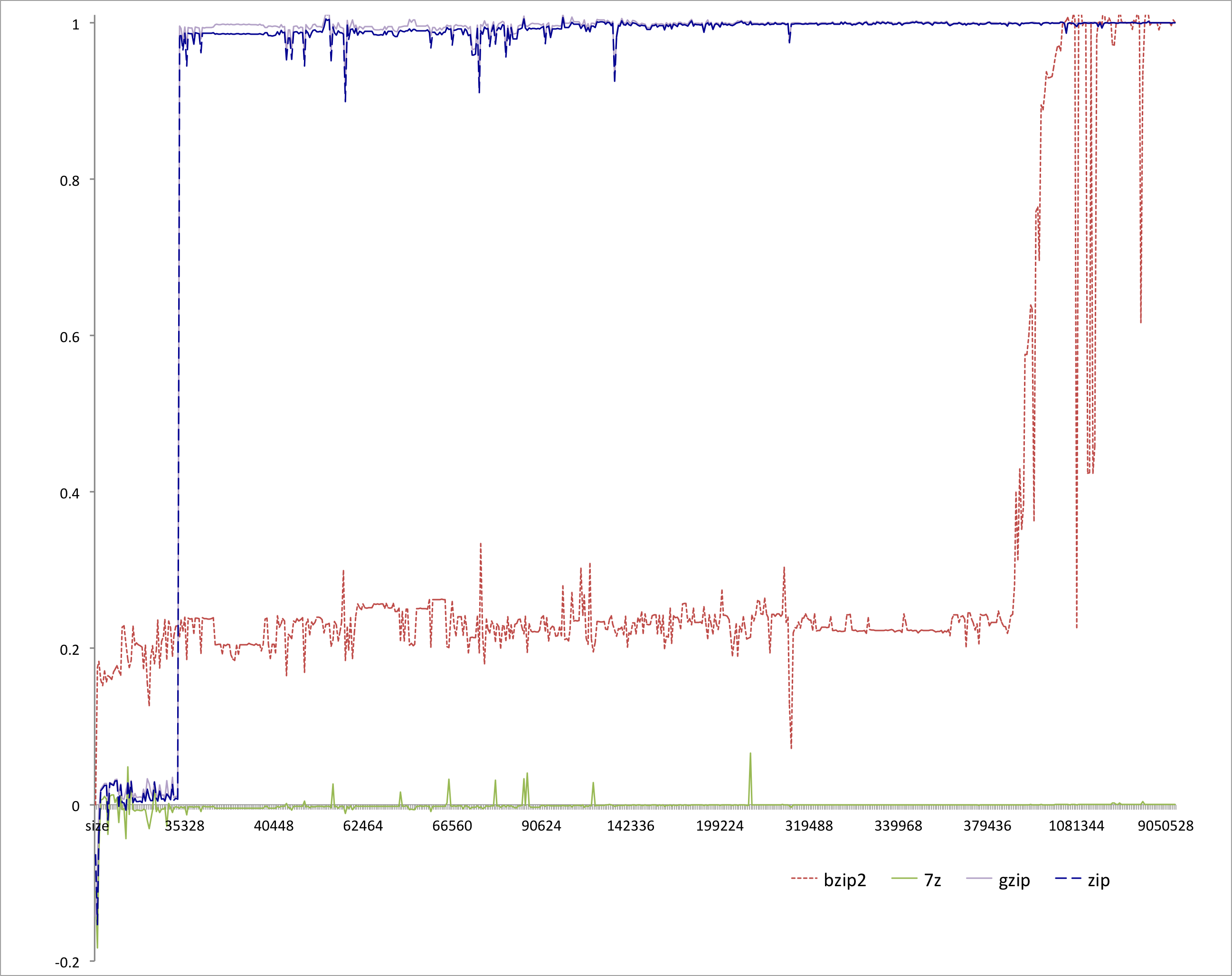}
  \end{center}
  \caption{The NCD of files to themselves for different file sizes using different compressors.}
  \label{fig:7zip}
\end{figure}

NCD is an upper bound on information distance. The choice of compressor
determines how tight this upper bound will be. Previous research
\cite{CAO+CIS2005} found that the size of strings we want to compare and
the size of the block or window that the compressor uses affect the values of
NCD. Our own experiments confirm this finding and indicate that a compressor
similar to 7-zip performs well (\ie using 7-zip, NCD(x,x) is close to zero) for
our domain, classifying malware and benign-ware.  \autoref{fig:7zip} shows the
results of our experiment. The $x$-axis is the size of the file while the
$y$-axis is the NCD of the file to itself. We can clearly see that 7-zip
outperforms the other three compressors (gzip, winzip and bzip2). The window
size in 7-zip can be set to a maximum of 4GB, making it suitable to calculate
NCD for two files with a combined size of up to 4GB.

\subsection{Classifier}

We build a classifier based on NCD to classify individual programs.
We design a random forest classifier using NCD and compressibility ratio
features to classify programs as benign or malicious. A decision
forest~\cite{Breiman2001} is an ensemble classifier composed of multiple, independent
decision trees. Trees are trained independently over a randomly selected 
subset of the features at each decision point.  When classifying a program,
each tree's decision is a vote; the forest's decision is affirmative when
the fraction of the affirmative votes of its trees exceeds a threshold. 
Tuning the decision threshold leads to a classifier with 
different true positive rate versus false positive rate trade-offs.  
Random forest training and classifying can be
parallelized, and they exhibit good classification and generalization
performance~\cite{CriminisiShotton2011}.

\vspace{2mm} \noindent {\bf Feature selection.}\ Our classifier takes as input
a vector of $n+1$ real-valued features in the range $[0.0, 1.0]$.  We form this
vector as follows.  The first feature is a program's \emph{compressibility
ratio}, the ratio of its compressed size to its uncompressed size. The other
$n$ features are the NCD of the program with $n$ other \emph{reference
programs}, $\frac{n}{2}$ benign ones and $\frac{n}{2}$ malicious ones. We
select these $n$ reference programs uniformly at random from our collection of
labeled programs.  We can show that some features are more important to the
forest's decision than others. In future work, we plan to extract, then exploit
this knowledge to choose features more effectively.  The binary features used
in the decisions in the decision trees in the random forest are random
thresholds on the one of the $n + 1$ features.

\vspace{2mm} \noindent {\bf Training.}\ To train the random forest classifier,
we start from a set of programs labeled as benign or malicious. The set of
training programs is disjoint from the reference programs used to compute the NCD
and from the evaluation set used subsequently to estimate the classification 
performance of our approach. 

Training a random forest involves training each decision tree independently,
while injecting sufficient randomization, as described below, to ensure robust
learning. Our approach is to use the full training set to train each forest,
but randomize the set of features made available to train each decision point
in each tree.  We defined this restricted feature set by picking an index into
the feature vector uniformly at random, then, to make the decision binary, we
pick a threshold, again uniformly, from all the vectors in the test set at this
index. 

We greedily built individual decision trees by selecting and storing a feature
at each decision point, out of the restricted set, to minimize the Shannon
Entropy over the labels of their leaves. Concretely, given an initial set of
items $B$ at a decision point $d$, we try to partition $B$ into $L$ over $d$'s
left subtree and $R$ over $d$'s right subtree to maximize the information gain:
\begin{align}
  I = H(B) - \frac{|L|}{|B|} H(L) + \frac{|R|}{|B|} H(R), \label{eq:I}
\end{align}
where $H$ denotes the Shannon entropy function over the distribution of labels in the set, and $|\cdot|$ the cardinality of the set.

Trees are grown to their maximum height, unless no proposed feature allows a significant information gain, set by a cut-off threshold. 
In each decision tree, each leaf stores the number of benign and malicious programs assigned to it.

The distribution of the training data plays a role in the accuracy of the
resulting classifier when applied to unknown items. 
The relative prevalence of labels in the training
set influences the maximization of information gain, as seen in \autoref{eq:I}. Increasing the prevalence of items with a
particular label in the training set introduces a higher penalty for
misclassifying those items. We therefore test our classifiers under different
training conditions to ensure they are robust, and detect good training
distributions to improve their performance.

Our training implementation is parallelized to use an arbitrary number of cores on a single computer, and can easily be ported to a distributed setting. 

\vspace{2mm} \noindent {\bf Classification.}\ Given an unknown program, we wish
to 
label it as benign or
malicious. First, its $n+1$ feature vector is computed by calculating its
compressibility ratio, and its NCD to the $n$ reference programs. Then each
decision tree in the forest 
assigns the program to a leaf: decisions based on the program's feature vector 
branch left or right, until a leaf is reached.
We interpret each leaf as an empirical probability the item is benign or
malicious, and the decisions of all trees are averaged to derive the overall
likelihood for each category. 

\vspace{2mm} \noindent {\bf Evaluation metrics.}\ The traditional metric for success for a classifier is accuracy, defined as the number of correctly classified items over the total number of items. Unfortunately, this measure confounds Type I and Type II errors.  Considering the classifier's false positive and false negative rates separately is easier to understand and interpret.

Since the output of a random forest is a real in $[0.0, 1.0]$, a decision boundary may be set within this range over which one classifies a program as malicious. Different decision thresholds exhibit different True Positive (TP) and False Positive (FP) rates. The Receiver Operating Characteristic illustrates the trade off between TP and FP for all possible decision boundaries. 

The concrete problem of malware detection is one where the base rate of positives (malware) may be significantly lower than 50\%. Thus, a key metric of success is the true positive rate (the fraction of positives classified as positives), for \emph{extremely low rates} of false positives (the fraction of negatives misclassified over all negatives). A high rate of false negatives would otherwise lead to the vast majority of items recognized as positives being misclassifications. For this reason, our evaluation illustrates ROC curves in the false positive range of 0\%--10\% only.


The choice of feature and training sets impacts the effectiveness of the classifier. Refining the selection process of those two sets and adding lightweight preprocessing steps (e.g., unpacking packed malware) is expected to improve results. In this paper, however, we want to isolate and investigate the performance of NCD and decision forests without the interference of other techniques.

Finally, some of the $n+1$ features may be more important than others' in providing information to classify programs as benign or malicious. To determine feature importance we follow the approach suggested by Breiman~\cite{Breiman2001} and report the fraction of trees that use a specific feature. We consider that a feature is used in a decision tree if any threshold was applied to it anywhere in a tree to decide the outcome of a branch.

\subsection{Lower Bound on NCD}
Computing NCD can be time consuming because it is a pairwise measure. The most expensive part of the computation is the compression of the concatenation of the two strings. Compressing each string separately is also time consuming but only needs to be performed once for each string while the compression of the concatenation needs to be performed for every possible combination of strings. 

We can improve the scalability of using NCD if we can find a way to minimise the number of comparisons we have to make. One way to achieve this is to approximate the lowest NCD we can obtain for a pair of strings. If this value is high (for example 0.8 or 0.9), then we know that the two strings are not similar and we can skip calculating their NCD. We now show how to compute such a lower bound, using only the compression of each string.  Assuming $Z$ is a normal compressor (\autoref{sec:background}) and that $Z(y) \le Z(x)$ (the case $Z(y) \le Z(y)$ is symmetric), we have 
\begin{align*}
e_Z(x,y) &= \frac{Z(xy) - \mbox{min}\{Z(x),Z(y)\}}{\mbox{max}\{Z(x),Z(y)\}} \\
	 &= \frac{Z(xy) - Z(y)}{Z(x)}.
\end{align*}

The smallest NCD value we can then obtain occurs when $x$, the longer string,
fully contains $y$, the shorter string.  In this case, $Z(xy) = Z(x)$ and we
have
\begin{align*}
e_Zmin(x,y) = \frac{Z(x) - Z(y)}{Z(x)}  = 1 - \frac {Z(y)} {Z(x)}.
\end{align*}
For example, if $Z(x) = 10$ and $Z(y) = 3$, we know, without compressing $xy$,
that the minimum NCD we can obtain is $0.7 = 1 - \frac{3}{10}$. We can then
choose a threshold, based on application domain, and compute $e_Zmin(x,y)$.
If $e_Zmin(x,y)$ exceeds this threshold, we are not interested in, and do
not compute, the exact $NCD(x,y)$.

\subsection{Evading NCD}

Large scale production of malware depends on the automated generation of variants of the same malware. To evade detection by NCD, variant generators must increase the NCD between the original malware and its variants.  To this end, they have two choices: add new content or obfuscate. 

If we apply the previous formula for NCD lower bounds to a file with itself, we have 
\begin{align*}
e_Zmin(x,x) =  1- \frac {Z(x)} {Z(x')}.
\end{align*}
We can clearly see that we need to add at least 100\% more unique content (increase $Z(x')$ by 100\%) to increase NCD to $0.5$. This NCD value might not even be enough to evade detection, as a $0.5$ similarity to known malware may still flag the new variant as suspicious to an NCD-based classifier.

Alternatively, we can evade detection via obfuscation or replacing content to increase $Z(xx')$, the compressed size of the malware $x$ concatenated to its variant $xx'$. Variant
generators must, as in the first case, obfuscate a large percentage of unique content to increase the NCD of $x$ and $x'$. This increase in the
effort needed to generate variants raises the bar for malware writers and makes
generating variants a more laborious and time consuming task. Adding high
entropy junk to each variant might overcome this obstacle. However, the large
increase in the size of the resulting variants is likely to undermine their 
viability by hindering their propagation. 

In our experiments, we found that the majority of false
positives occurred in setup and installer files. These files perform actions that, from first principles, are similar to malware actions (e.g., change the file system or the registry). This result suggests that NCD might
be able to capture behaviour typical of malware even across different malware families.

\section{Evaluation}
\label{sec:eval}

We designed our study to answer the following research questions:
\begin{itemize}[]

\item[RQ1:] \emph{How accurate is NCD in classifying malware?}
 
 As mentioned before, NCD is lightweight in that it can be applied directly to the binary executables without the need for reverse engineering or executing the programs. This research question investigates NCD's performance (accuracy and false and true positive rates) in classifying programs as malware or benign-ware.  
 
\item[RQ2:] \emph{How accurate is using compressibility rates in classifying malware?}

We expect malware to have higher entropy than benign-ware because, in the current state of the malware arms race, malware designers often use polymorphism (compression and encryption) to avoid detection. Therefore, we expect malware to be less compressible than benign-ware. Using compressibility rates is less computationally expensive than NCD because compressibility rate is a feature of an individual program while NCD is a pairwise feature. In this question we investigate the performance of compressibility rates (accuracy and false and true positive rates) compared to NCD in classifying malware. 

\item[RQ3:] \emph{Is malware reported in the same timeframe more likely to have similar patterns?}

The malware we originally collected was reported by the public on the VirusWatch Archive in a period of 8 consecutive days. This fact might influence results if malware reported within a short time period is more homogenous. We investigate this issue by repeating our experiments on a new sample set that was reported on randomly selected dates spread throughout a little under two years and examining the results.

\item[RQ4:] \emph{How much can we reduce the cost of using NCD by using approximations based on NCD lower bounds?}

The performance of an NCD based classifier is expected to rely on the number of programs that are used to build and train the classifier. However, a larger number of samples leads to a larger number of NCD pairwise comparisons. Since the time complexity for NCD is quadratic, the effect of increasing the sample set on performance might make the approach impractical. We can use NCD lower bounds to approximate NCD and reduce the cost. 
However, before using approximations in the classifier, we first empirically quantified savings, in the number of computations, this approximation achieves in practice.   

\item[RQ5:] How does an NCD classifier compare to commercial and open-source anti-virus software?

We labeled the programs in our samples as malware or benign-ware based on the source from which we obtained them. Although we expect this classification to be reasonably accurate, we can not guarantee that it is 100\% accurate as benign-ware might be mistakingly reported as malware to the virus repository we used. Similarly, we cannot guarantee that no malware exists in the benign-ware set. To have a ground truth, we need to reverse engineer and analysis each program in the sample set; a labour intensive solution that is not practical. However, we can leverage anti-virus software to gain a better understanding of our samples and our results. 

\end{itemize}

\subsection{Corpus}
\label{sec:corp}

We collected malware and benign-ware to form the corpus for our experiments in the following way: For malware, we used a script to automatically download all executable binaries that were reported on the VirusWatch Archive\footnote{\url{http://lists.clean-mx.com/pipermail/viruswatch/}} from the 4th until the 11th of April. We configured the script to attempt to download a file a maximum of 3 times and to abort a connection after 5 seconds of idle time. 

For benign-ware, we collected all executables from two Windows 7 machines. We found collecting benign programs to be more challenging than collecting malware because benign-ware sources do not offer executables but rather offer installers and setup files. Collecting only installer files would not be representative of benign-ware, while using them to install applications and then collect the resulting binary files proved to be a laborious. Therefore, we decided to collect executables from existing machines. 

After collecting all samples, we filtered the results to retain only Windows executables using the results of the Linux \texttt{File} utility. We then used the Linux \texttt{fdupes} utility to identify and remove any duplicate files. The \texttt{fdupes} utility compares files first by size, then MD5 signature and finally uses a byte-by-byte comparison. Table \ref{tab:stats} provides some statistics about our corpus.

\begin{table}[t]
\centering
\setlength{\tabcolsep}{1pt}
\begin{tabular}{ lrrrrrrrrr}
\toprule
       &   \multicolumn{3}{c}{All Samples}   &  \multicolumn{2}{c}{Set 1} &  \multicolumn{2}{c}{Set 2}&  \multicolumn{2}{c}{Set 3} \\ 
                           &   &  \multicolumn{2}{c}{Size (KB)}   &  \multicolumn{2}{c}{Size (KB)} &  \multicolumn{2}{c}{Size (KB)}&  \multicolumn{2}{c}{Size (KB)} \\ 
      Type                      & Num   &  mean & med.  &  mean & med.&  mean & med.&  mean & med.  \\ 
\midrule
       Benign                &     3,046       &  932          &  116   &903 &118 &766 & 112  & 1,006 & 110 \\
       Malware      &         14,656        &    4,287        &   500  &  3,120& 500 & 3,279  & 501 & 2,449 & 501   \\
\bottomrule
       \end{tabular}
\caption{Descriptive statistics of the malware and benign samples used in our study.}
     \label{tab:stats}
\end{table}

To conduct our experiments and have multiple data points that can allow for statistical analysis, we randomly sampled 1,000 programs from each type (malware and benign-ware) from our corpus. We repeated the sampling process (with replacement) three times creating three sets of 2,000 programs each. We then conducted each experiment in our study on each sample set independently and then analysed and compared results. Statistics about the the samples sets can also be found in Table \ref{tab:stats}.

\subsection{Classifier Parametrisation}
\label{sec:classparams}

The Random forest classifier we built to label programs as benign or malicious has a number of parameters that need to be fixed before training and classification may take place. 

The random forests we trained for all experiments consist of 400 individual decision trees trained independently in parallel. Randomness is injected into the training process by considering a fresh random selection of 30 features (out of the available $n+1$) for training each tree branch, but making all training data available. Trees are grown, adding branches, until the information gain is less than $0.001$ bits, or a depth of $5$ branches has been reached.

We experimented with training under two conditions: first, we train the classifier with an equal number of benign and malicious training exampled; second, we severely bias the training set by using 10\% benign and 90\% malicious examples. The biased training condition heavily penalizes false positives, and should lead to fewer of those at the expense of increased false negatives.

\subsection{NCD Classifier}
\label{sec:ncd:classifier}

\begin{figure}[t]
  \begin{center}
    \includegraphics[width=0.95\columnwidth]{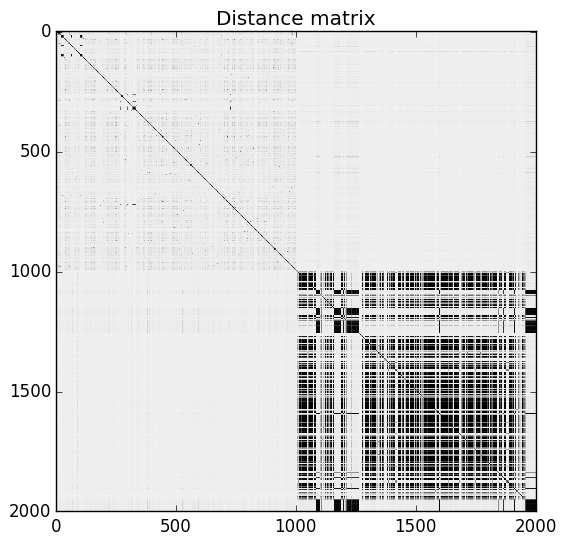}
  \end{center}
  \caption{NCD matrix for Sample set 1.}
  \label{fig:distances1}
\end{figure}

For each of our three 2,000-program sample sets described in Section \ref{sec:corp} , we first calculated pairwise NCD using the 7-zip compressor. Figure \ref{fig:distances1} is the distance matrix for sample set 1: the darker dots represent similarity (lower NCD) while lighter areas represent dissimilarity (higher NCD). The programs are ordered in the matrix such that benign-ware comes first followed by malware. The matrix shows that the malware in the set is clearly homogenous while benign-ware is similar to neither other benign-ware nor malware. Similar results are observed for the two other sample sets.

We applied the algorithm described in \autoref{sec:approach} to the three sample sets. We set the number of features to $n = 200 + 1$, namely the compression ratio and the NCD from 200 randomly chosen programs --- 100 known benign and 100 known malicious. 
The size of the training and test sets as 600 and 600 respectively (with 300 benign and malicious sample each). 
Because the results are dependent on the choice of sub sets, we repeated the experiment 30 times for each set by randomly selecting different subsets to act as features, training and test sets. We note that the size of the sample gives limited resolution for very small True Positive Rates and False Positive Rates. This limitation can be overcome by using larger (but more difficult to gather) corpus of malware.

\begin{figure}[t]
  \begin{center}
    \includegraphics[width=0.95\columnwidth]{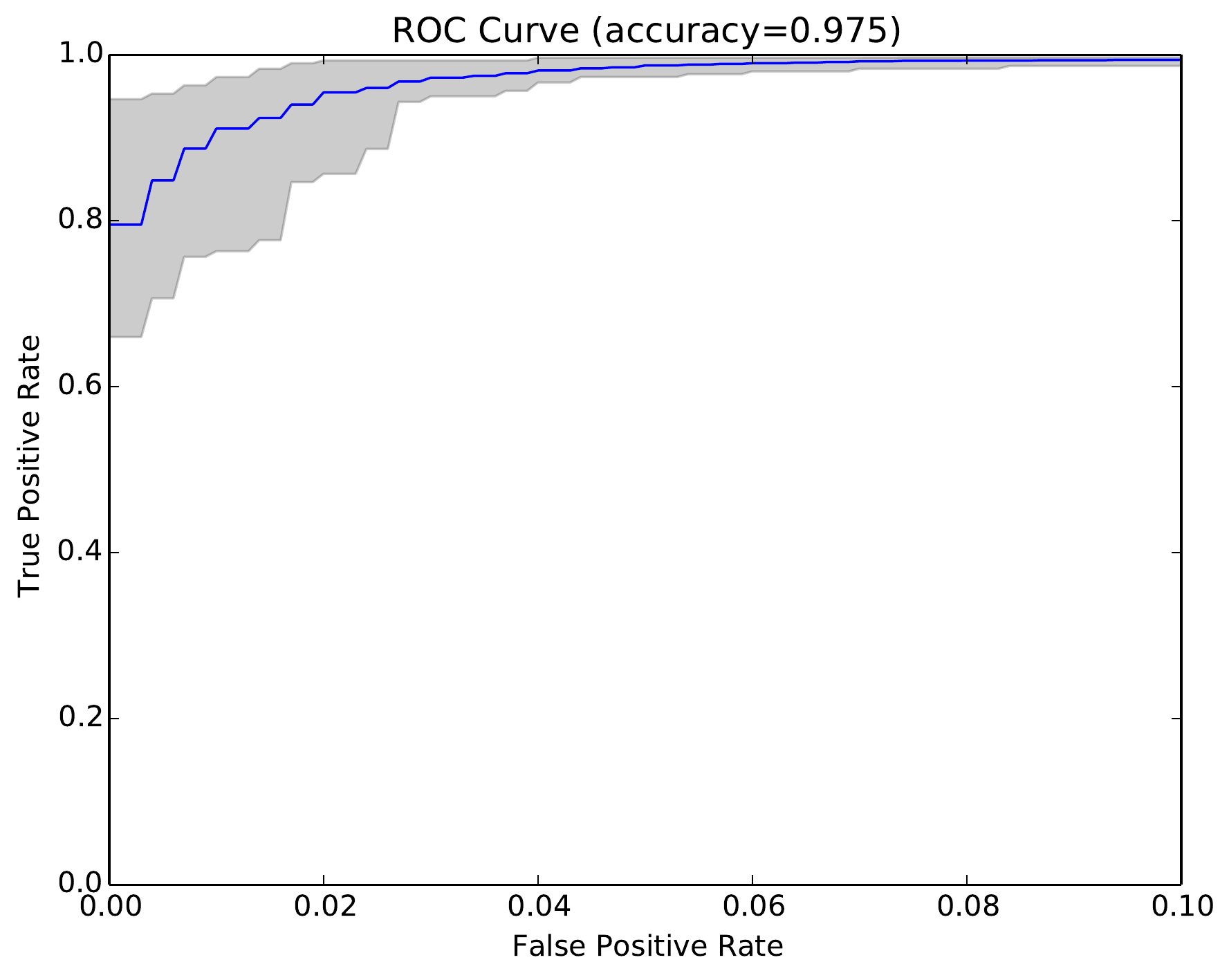}
  \end{center}
  \caption{ROC curve for 30 runs of the NCD classifier for Sample set 1.}
  \label{fig:roc1}
\end{figure}

The results for each run are represented as a Receiver Operating Characteristic (ROC) curve. \autoref{fig:roc1} shows the ROC curve for the 30 runs of the experiment for sample set 1. The blue curve shows the average results while the grey area is between the maximum and minimum of the 30 runs.  Each point on the curve represents the accuracy rates we can achieve by using a different vote threshold to differentiate between malware and benign-ware.
 
The x-axis is the false positive rate while the y-axis is the true positive rate.  The maximum accuracy that can be achieved across all runs (displayed at top of the graph) is 97.5\%. The average maximum accuracy achieved is 97.1\% with an average false positive rate of 3\% and average true positive rate of 97.3\%. However, if we choose a different threshold, we can reduce false positives or increase true positives by sacrificing accuracy. The choice of threshold depends on the objective we want to achieve. For example, we can have no false positives if we chose a more conservative threshold, but accuracy and true positive rate would be reduced, on average, to 92.2\% and 84.9\% respectively. On the other hand, we can set the threshold higher and capture all true positives but also capture on average around 22\% false positives. 

The results for the other two samples are similar. The first set of columns of Table \ref{tab:ncdres} (under header NCD) shows the average false positive (FP), true positive (TP) and accuracy rates (Acc) for our experiments over 30 runs for each of the sample sets. 

\begin{table}[t]
\centering
\setlength{\tabcolsep}{1pt}
   \begin{tabular}{ lrrrrrrrrr}
\toprule
   &   \multicolumn{3}{c}{NCD}   &  \multicolumn{3}{c}{Comp Rate} &  \multicolumn{3}{c}{Combined} \\ 
   Sample & FP  &  TP  &  Acc &  FP  &  TP  &  Acc & FP  &  TP  &  Acc \\ 
\midrule 
      Set 1            & 0.030  & 0.973  & 0.971 & 0.057  & 0.961& 0.952  & 0.030 & 0.974  &0.972 \\ 
      Set 2            & 0.034  &  0.978 & 0.972 & 0.037 & 0.961& 0.962  &0.030 & 0.985  & 0.977 \\ 
      Set 3      &  0.034 & 0.976  & 0.971 & 0.060 &0.969 & 0.955  & 0.030&  0.978 & 0.974\\  
\midrule
      All          &  0.032 & 0.976 &0.971 & 0.051 &0.964 & 0.956& 0.030 & 0.979 & 0.974 \\ 
\bottomrule
       \end{tabular}
\caption{Average best value for accuracy with average corresponding false positive and true positive rates for each sample set and each approach.}
\label{tab:ncdres}
     
  \end{table}

 To understand how NCD and decision forests contributed to the observed results, we used a simple clustering algorithm  (k-medoids) to cluster the programs in each sample set using NCD as the distance. In k-medoids clustering, {\em k} random points are selected as the centres (or medoids) of each cluster. All other data points are then assigned to the cluster in which they are closest to the medoid. A random point in the cluster is then swapped with the medoid and the calculations are recomputed. If the swap causes the cost to be reduced (the sum of distances within each cluster), the new medoid is used, otherwise the original medoid is kept. The process is repeated until there is no change in medoids. We applied this algorithm to each sample set setting {\em k} to 35. We labeled each cluster as a malware or benign-ware cluster based on the dominating number of samples in the cluster. We then calculated false and true positive rates as well as accuracy. The results (Table \ref{tab:ncdclust}) show that NCD alone can achieve high levels of accuracy in classifying malware and benign-ware (95.6-95.8\%), however using decision forests improves performance.

\begin{table}[t]
\centering

\begin{tabular}{ lrrr}
	\toprule   
	Sample                & FP  &  TP  &  Acc  \\ 
	\midrule
	Set 1            & 0.052 & 0.968  & 0.958  \\ 
	Set 2            & 0.040 & 0.952  & 0.956 \\ 
	Set 3      &  0.042 & 0.953  & 0.956 \\  
	\midrule
	All        & 0.045 & 0.958 & 0.957 \\ 
	\bottomrule
\end{tabular}
\caption{Accuracy, false positive and true positive rates for each sample set using NCD and k-medoids clustering with 35 clusters.}
\label{tab:ncdclust}
\end{table}    
  
 {\bf The answer to RQ1 is} that NCD is effective in classifying malware and benign-ware and when used in conjunction with decision trees can achieve an average accuracy of 97.1\%  with average false positive rates of 3.2s\% and true positive rates of 97.6\%.    
  
  \subsection{Compressibility Rate Classifier}
  
\begin{figure}[t]
 \begin{center}
   \includegraphics[width=0.95\columnwidth]{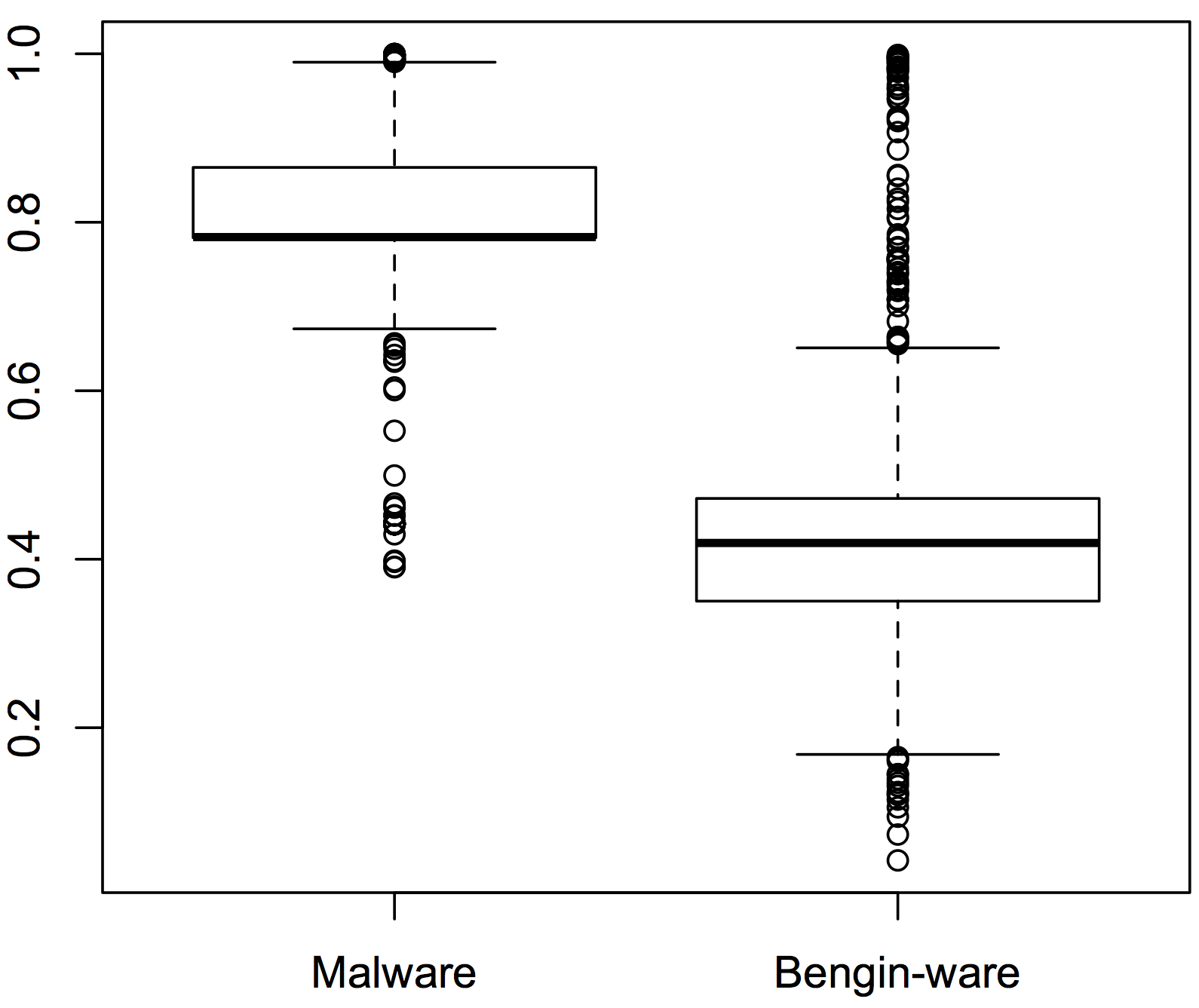}
 \end{center}   
  
  \caption{Compressibility rates of malware and benign-ware for sample set 1.}
  \label{fig:comprates}
\end{figure}

 As mentioned before, we expect malware and benign-ware to have different compressibility rates ({\em compressed size $/$ original size}). We plotted the compressibility rates of our samples to test our intuition. Figure \ref{fig:comprates} shows the box plots for sample set 1: The boxes represent the middle 50\% compressibility rates of the sample, divided by the line that represents the median while the whiskers represent the top (and bottom) 25\%.  The box plots confirm that there is a clear difference in compressibility patterns between malware and benign-ware. Similar results are observed for the other two sets.

\begin{figure}[t]
  \begin{center}
    \includegraphics[width=0.95\columnwidth]{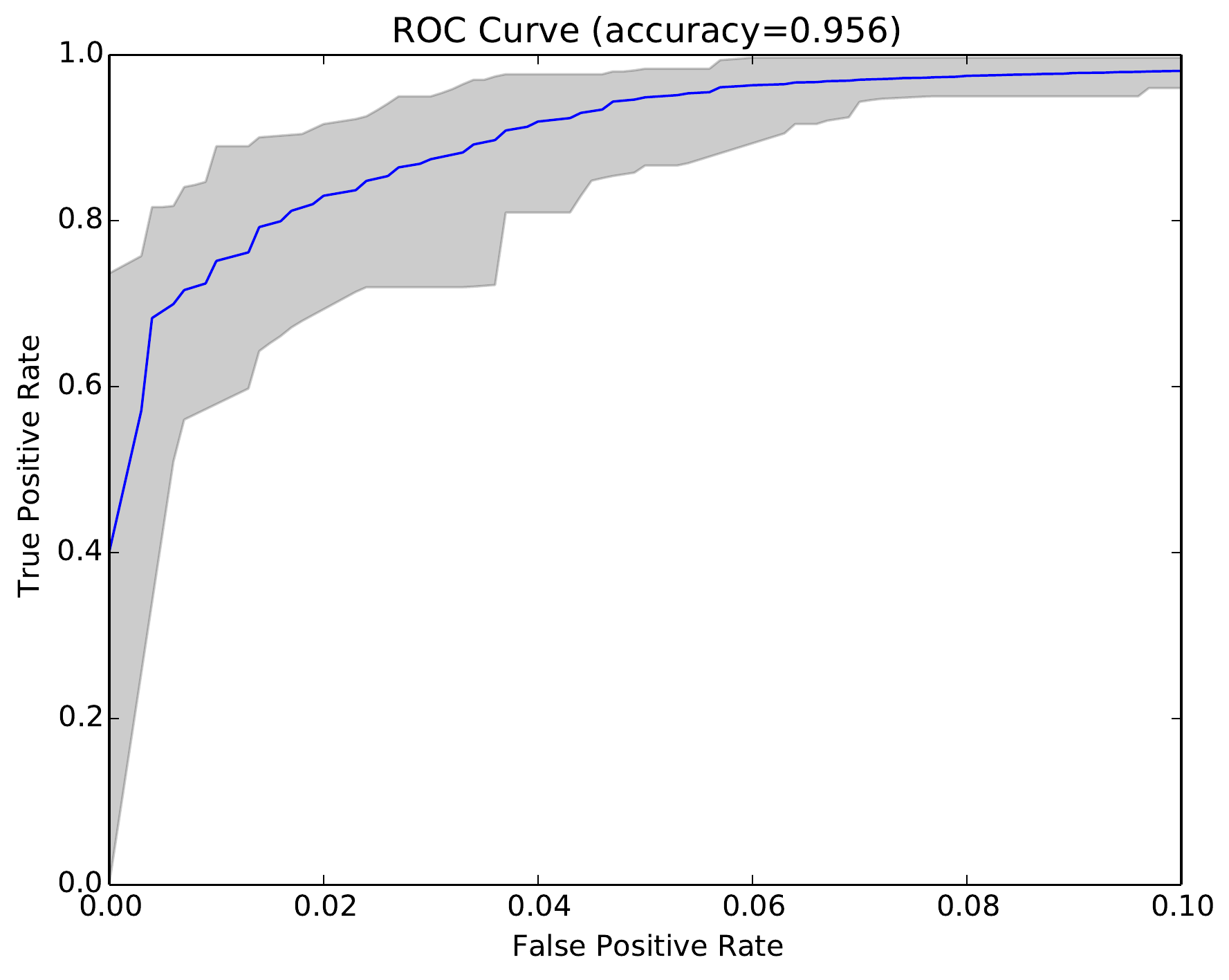}
  \end{center}
  \caption{ROC curve for compressibility rate classifier over 30 runs for Sample set 1.}
  \label{fig:roc2}
\end{figure}

Inspired by these encouraging results, we repeated the experiments in {\bf RQ1} using compressibility rates as features instead of NCD. The main benefit of using compressibility rates (CR) instead of NCD is that NCD is a pairwise calculation while CR is just calculated once for each program. The results (second set of columns in Table \ref{tab:ncdres}) show that although the accuracy achieved with CR is lower than that obtained by NCD, the reduction is on average around 1.5\% (97.1\%--95.6\%). Figure \ref{fig:roc2} shows the ROC curve (blue curve) and variation (grey area) in false positive and true positive rates using a compressibility rates classifier using the same sample of programs as Figure \ref{fig:roc1}. 
 
\begin{figure}[t]
  \begin{center}
    \includegraphics[width=0.95\columnwidth]{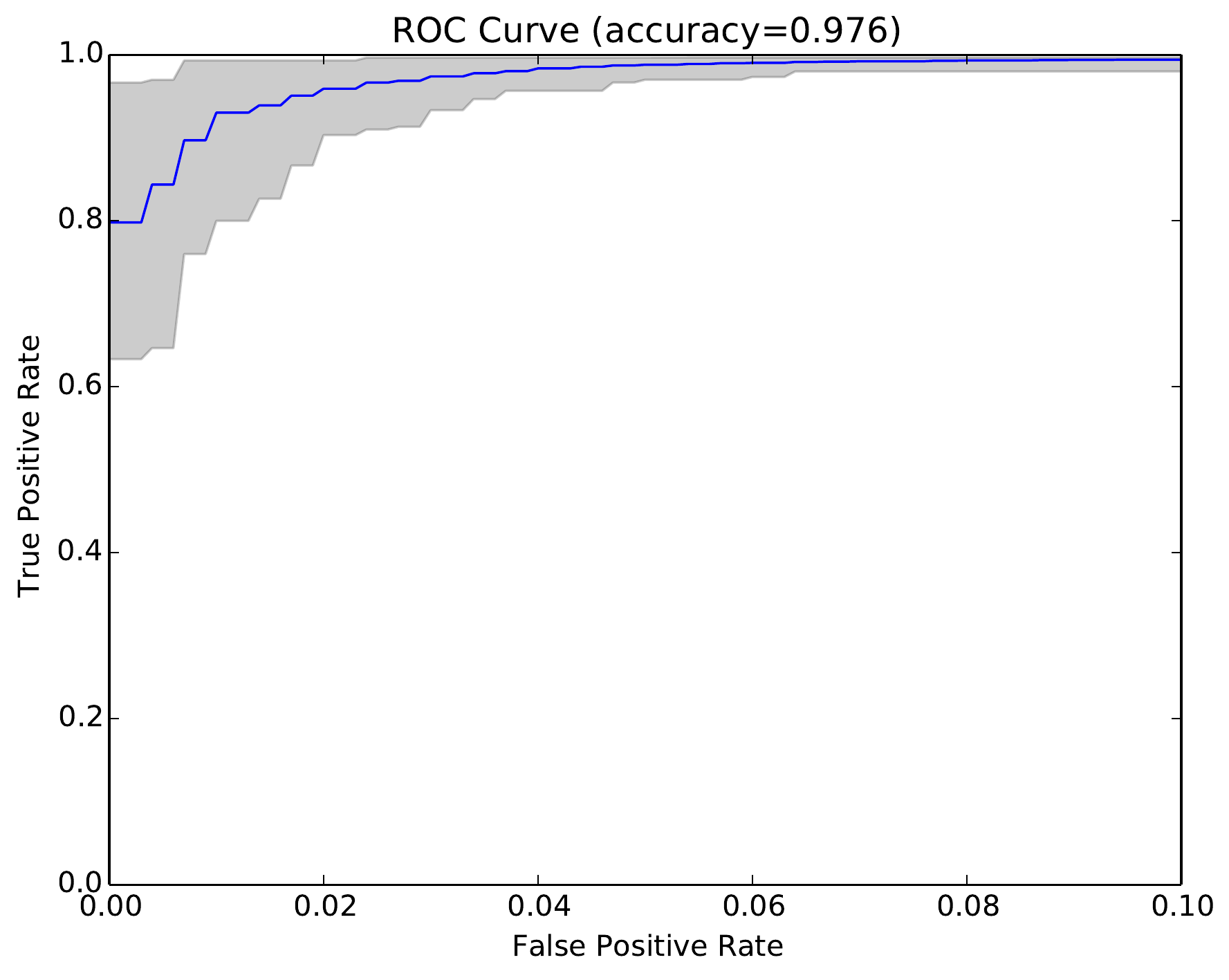}
  \end{center}
  \caption{ROC curve for 30 runs of the combined compressibility rate and NCD classifier for Sample set 1.}
  \label{fig:roc3}
\end{figure}

Finally, we tried combining NCD with compressibility rates to see if the two approaches are complementary. The combined approach achieves on average higher accuracy (97.4\% vs 97.1\%), lower false positive rates (3\% vs 3.2\%) and higher true positive rates (97.9\% vs 97.6\%). The average results over 30 runs are reported in the last set of columns in Table \ref{tab:ncdres}) while Figure \ref{fig:roc3} shows the roc curve for the same sample as Figures \ref{fig:roc1} and \ref{fig:roc2}.  
 
We applied a two-sided Mann-Whitney test to the accuracy observations and found that the differences between the three approaches are statistically significant with 95\% confidence.

 {\bf The answer to RQ2 is} that compressibility rates are effective in classifying malware and benign-ware and can achieve an average maximum accuracy of 95.6\%  with average false positive rates of 5.1\% and true positive rates of 96.4\%. NCD classifiers perform statistically significantly better than compressibility rate classifiers. Combining NCD and compressibility rates statistically significantly improves performance.

\subsection{Size of Malware Reporting Window}

A threat to validity of our study is that the malware we collected was reported within a short time-frame. If malware spreads in outbreaks where several variants are released by an adversary at the same time, this might suggest that our approach could lose effectiveness over time. To investigate this issue, we repeated all our experiments with samples that have more diverse reporting times. We selected 10 random dates from the period between 1/1/2013 and 1/10/2014. We then downloaded all the Windows executable malware reported on those 10 dates and processed them in a similar manner to our original corpus. We then randomly selected 1,000 malware from this new corpus and 1,000 benign programs from our original set of benign-ware and repeated the experiments in RQ1 and RQ2.
  
\begin{figure}[t]
  \begin{center}
    \includegraphics[width=0.95\columnwidth]{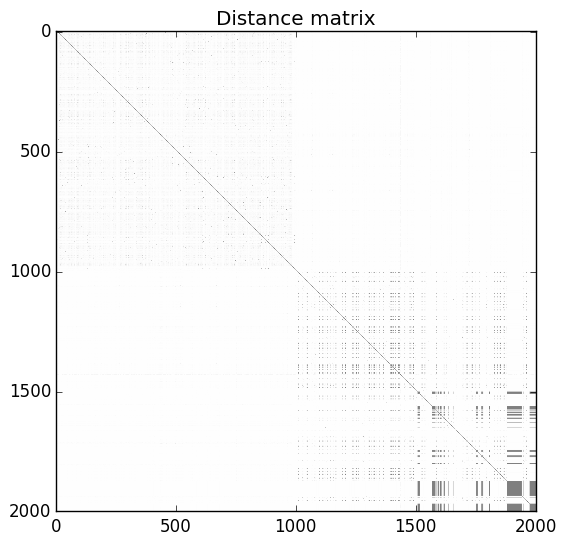}
  \end{center}
  \caption{NCD matrix for a diverse (with regards to reporting time) sample set.}
  \label{fig:distances4}
\end{figure}

Figure \ref{fig:distances4} shows the distance matrix for the new diverse set. The malware in this sample is noticeably less homogenous than the previous sets. However, the same general conclusion still holds: malware is more similar to malware than benign-ware while benign-ware is not similar to other benign-ware or malware. 
 
\begin{figure}[t]
  \begin{center}
    \includegraphics[width=0.95\columnwidth]{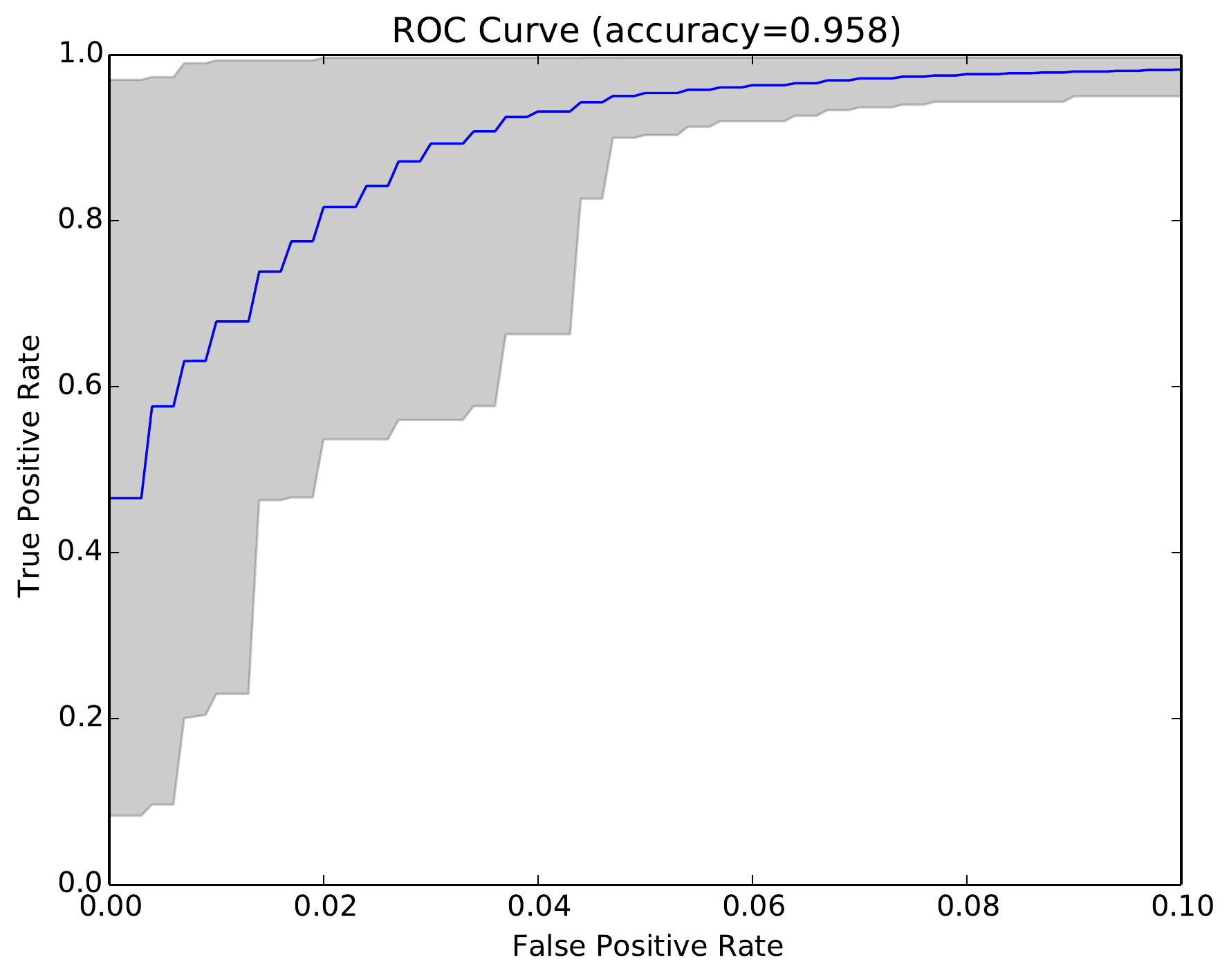}
  \end{center}
  \caption{ROC curve for 30 runs of the NCD classifier for the diverse sample set.}
  \label{fig:roc4}
\end{figure}

\begin{figure}[t]
  \begin{center}
    \includegraphics[width=0.95\columnwidth]{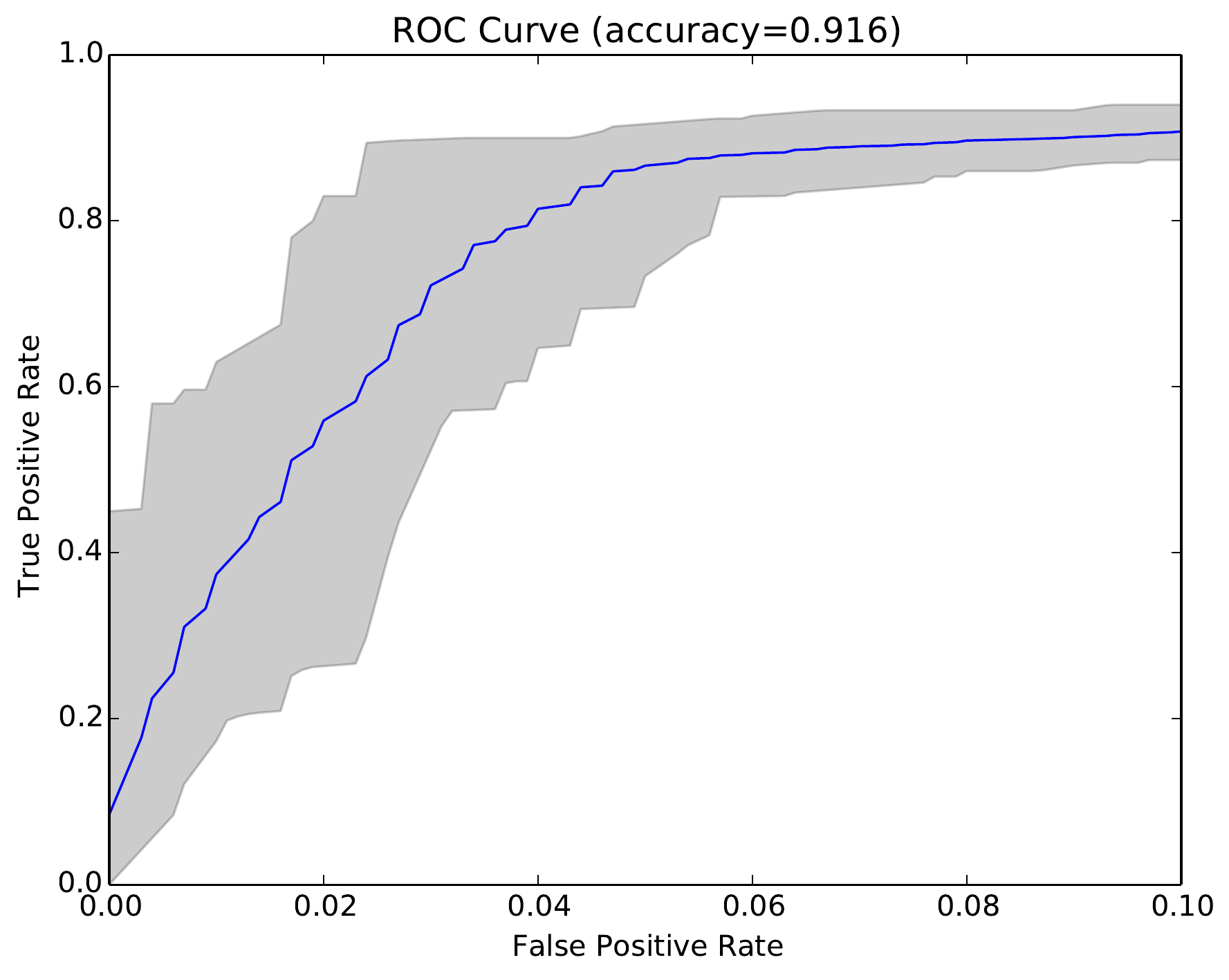}
  \end{center}
  \caption{ROC curve for 30 runs of the compressibility rate classifier for the diverse sample set.}
  \label{fig:roc5}
\end{figure}

\begin{figure}[t]
  \begin{center}
    \includegraphics[width=0.95\columnwidth]{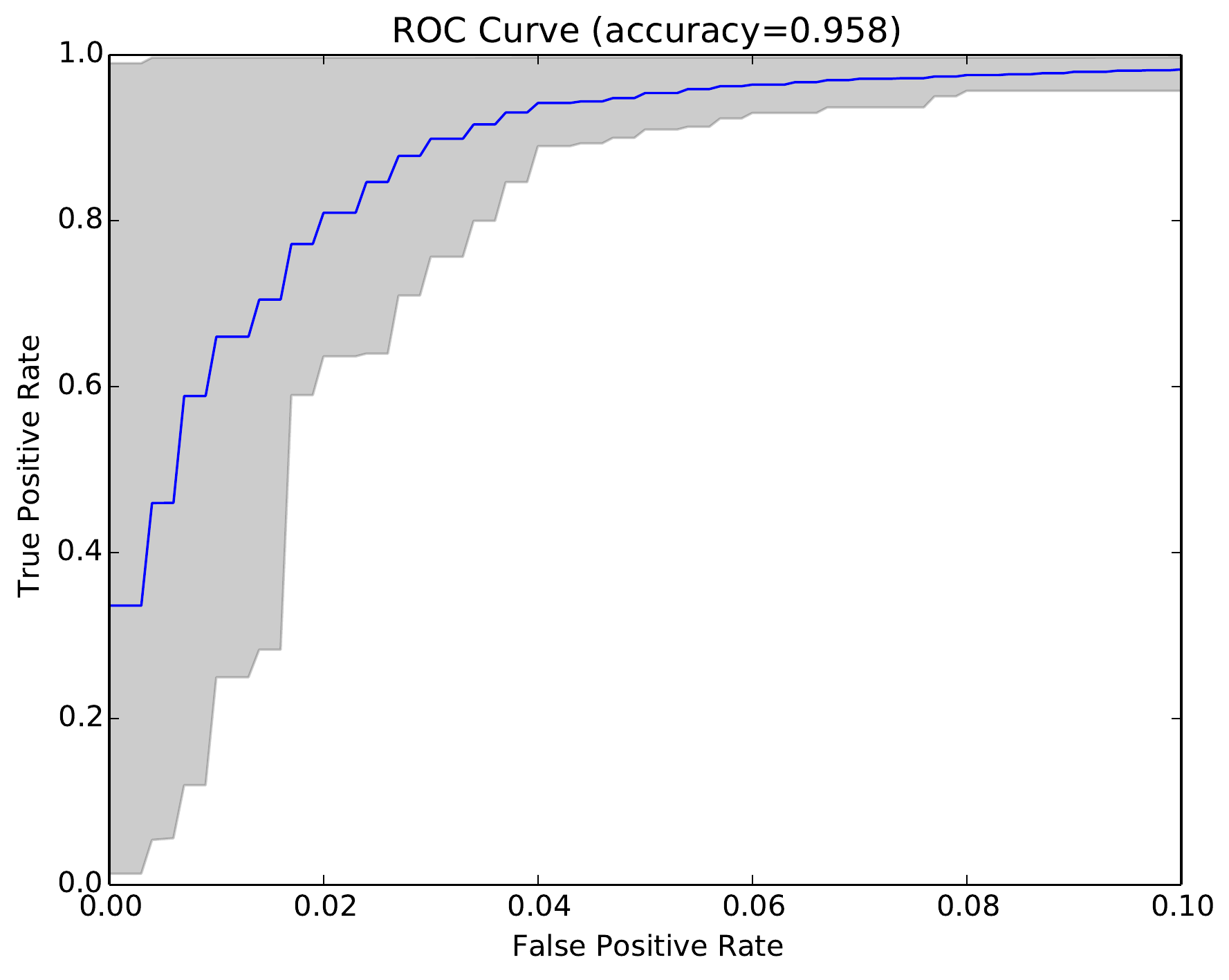}
  \end{center}
  \caption{ROC curve for 30 runs of the combined compressibility rate and NCD classifier for the diverse sample set.}
  \label{fig:roc6}
\end{figure}

\autoref{tab:divres} depicts the average best accuracy across 30 runs of the classifier for NCD, compressibility rates and the combined approach. Figures \ref{fig:roc4}, \ref{fig:roc5} and \ref{fig:roc6} show the ROC curves for our 30 runs for the NCD classifier, the compressibility rate classifier and the combined classifier. The overall average accuracy results for each approach are slightly lower than those obtained from the previous samples. Interestingly, the results in these figures contain higher variation, as you can observe in the larger grey area in each graph, compared to Figures \ref{fig:roc1}, \ref{fig:roc2} and \ref{fig:roc3}.  This higher variance is expected; it is a consequence of the greater malware diversity of this data set.  Further, this variance and 
the tailtale staircase pattern of low resolution in the ROC curves is confined to FPR below $5$\% (due to the size of our sample sets, \autoref{sec:ncd:classifier}) but disappears above that threshold.  

Inference always depends on the choice of feature and training sets;  this result suggests that, for our approach to be successful in practice, we may need to evolve the classifier over time. This might not be a significant concern since the process of building the classifier is completely automated and safe (since it does not require running the malware).  This result also suggests that rebuilding the classifier need not be frequent since, even across a two-year interval, the classifier is still reasonably accurate. 

\begin{figure}[t]
 \begin{center}
   \includegraphics[width=0.95\columnwidth]{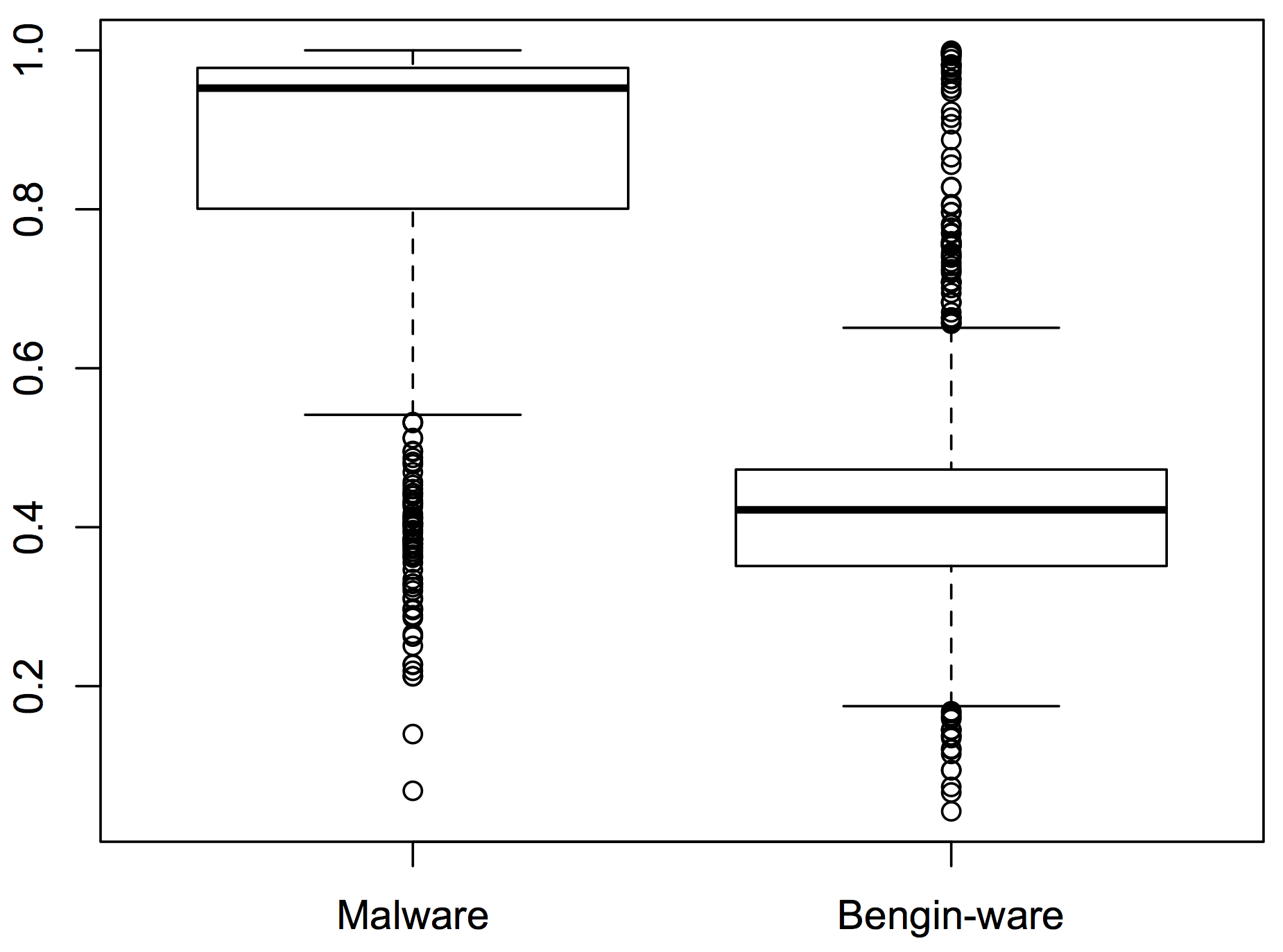}
 \end{center}   
  
  \caption{Compressibility rates of malware and benign-ware for  the diverse sample set.}
  \label{fig:comprates4}
\end{figure}

\autoref{fig:comprates4} shows the difference in compressibility rates between malware and benign-ware for the diverse set. Similar to the previous sets, there is a clear difference between the two sets. However, the average accuracy is much lower than the accuracy observed for the previous sets. This might be caused by the larger number of outliers in malware that have a lower compressibility rate. \autoref{fig:importance}, which is representative of the importance gain of our experimental runs, suggests that our data contains ``Kevin Bacon'' programs that effectively partition the rest of the programs because they lie near, and even define, the centre of clusters in the data.  In the upper left of the \autoref{fig:importance}, these programs give rise to the NCD measures that are the most prevalent feature. Thus, another reason for the drop in our accuracy may be due to the fact that the more diverse a data set is, the less likely one is to select these highly discriminant, ``Kevin Bacons''.

{\bf The answer to RQ3 is} that malware reported in a tighter timeframe is more
homogeneous than malware reported over a longer period of time.  Nonetheless,
our NCD classifier still achieves 95.2\% accuracy with 5\% false positive and
95.4\% true positive rates, on average.

\begin{table}[t]
\centering

   \begin{tabular}{ lrrr}
        \toprule    
      Approach       & FP  &  TP  &  Acc  \\ 
	\midrule
      NCD            & 0.050 & 0.954 & 0.952  \\ 
      Comp Rate            & 0.057 & 0.879  & 0.911 \\ 
      Combined     &  0.057 & 0.962  & 0.953 \\  
	\bottomrule
       \end{tabular}
   \caption{Average best value for accuracy with average corresponding false positive and true positive rates for the diverse sample set and each approach.}
     \label{tab:divres}
     
  \end{table}    
  
\subsection{NCD Cost Reduction}
As mentioned before, we can reduce the cost of computing NCD in terms of time and computation power by skipping calculations that we think might not contribute to the effectiveness of the classifier . For example, we might decide that if the lower bound on NCD for a pair of programs is $0.99$, the effort needed to compute the real value (which would be between $0.99$ and $1$) is not justified.

\begin{figure}[t]
 \begin{center}
   \includegraphics[width=0.95\columnwidth]{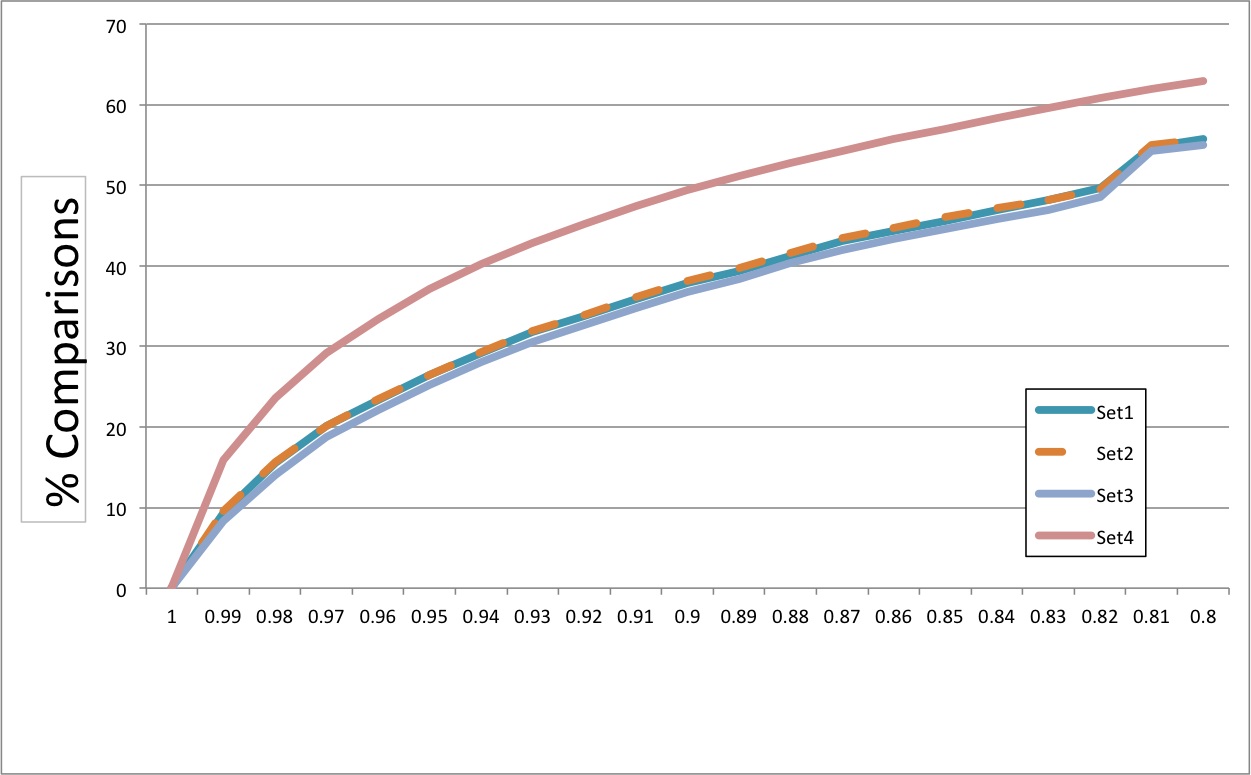}
 \end{center}   
  
  \caption{Savings in percentage of NCD calculations using lower bound  thresholds.}
  \label{fig:lower}
\end{figure}

We empirically estimate the percentage of calculations that we can skip by using lower bounds on NCD by setting the threshold at different values. For each of our four sets, we count the number of pairs where the lower bound on NCD is equal to or above a number of thresholds that range from $0.8$ to $1$ with $0.01$ increments. \autoref{fig:lower} shows the plot for each set. The x-axis is the different thresholds of NCD while the y-axis is the saving in percentage of computations. Each set consists of 2,000 programs; therefore, the total number of computations needed is $2,001,000$.    

The graph shows that the results for the three original sets is almost identical while the diverse set offers even more savings at each point of the threshold. If we set the NCD lower bound threshold at which we rely on approximation to $0.99$, we already save 8--16\% in the number of calculations. The percentage of calculations we can save grows gradually with the threshold and reaches between 37 to 49\% even at a threshold as high as $0.9$. These results indicate that we can reduce the cost of using NCD by using this simple technique of approximating lower bounds. Approximating NCD lower bounds can be considered a more refined way to use compressibility rates to classify, since these lower bounds compare the compressed size of two strings individually rather than the compression of their concatenation.

{\bf The answer to RQ4 is} that we can achieve savings from 8--16\% by setting the lower bound threshold to use approximations to $0.99$. The savings we can achieve increase as we reduce the threshold and reach 55--63\% for a $0.8$ threshold. 
   
\subsection{Comparison to Anti-Virus Software}

To gain a better understanding of our results we scanned all samples using Virus Total. Virus Total \footnote{\url{https://www.virustotal.com}} (a subsidiary of Google) provides an online service that scans files using up to 59 different anti-virus engines. The service also provides an API that can be used in batch scanning. We used uirusu \footnote{\url{https://github.com/arxopia/uirusu}} which is an interface written in Ruby to simplify uploading and scanning files using the Virus Total API. 

The Virus Total website only accepts files smaller in size than 64MB while the API only accepts files smaller than 32MB. It is sufficient to submit the MD5 hash code of a program if the file itself was submitted and analysed previously by another user. We scanned our malware and benign-ware samples using the API. 
Naturally, because we have no ground truth, we can only report on false positives and negatives by assuming that our initial labelling is correct. However, the results of scanning might give us an insight into our results and our samples.
  
\begin{table}[t]
\centering
   \begin{tabular}{ lrrrr}
        \toprule                   
	& \multicolumn{2}{c}{Set 1}   & \multicolumn{2}{c}{Diverse Set}  \\ 
        & Mal. & Ben. & Mal. & Ben. \\ 
\midrule
        Scanned & 994 & 1,000 & 980 & 1,000 \\
      Detected           & 941 & 94 & 789 & 105 \\ 
      Not Detected            & 59 & 906  & 206 & 895\\ 
      Anti-Virus Highest     &  883 & 45  & 525 & 47\\ 
       Anti-Virus Lowest     &  0 & 0  & 0 & 0 \\  
\bottomrule
       \end{tabular}
\caption{Summary of results of scanning sample set 1 and the diverse set using Google's Virus Total service.}
     \label{tab:vtsum}
\end{table}

Table \ref{tab:vtsum} summarises the results of the Virus Total scan on sample set 1 and the diverse set (sets 2 and 3 show similar results to set 1).  The first row shows the number of scanned files. All programs were scanned successfully except $6$ malware programs from sample set 1 and $20$ from the diverse set, which violated the size limitation. The number of programs classified by at least one anti-virus engine as malware is shown on the second row. Interestingly, the number of malware detected in sample set 1 is considerably higher than the number detected for the diverse set (941 vs 789). This mirrors our NCD classifier results where average best accuracy across 30 runs for sample set 1 was higher than that for the diverse set (97.1\% vs 95.2\%). This result might indicate that the difference in NCD classifier accuracy for the two sets is caused by inaccuracy in the original labelling rather than the diversity of the sample set.  However, more experiments have to be conducted to verify this observation. 

We also notice that in both sets, around 10\% (94 and 105) of benign-ware was classified by at least one anti-virus engine as malware. Examining the names of these misclassified programs reveals that they are mostly uninstallers, setup files and updaters. Interestingly, we observed a similar phenomenon in our NCD classifier. The similarity between malware and uninstallers, that was detected by NCD, might be caused by the fact that they perform similar actions (e.g., writing to the file system, changing the registry). This might suggest that some of the anti-virus engines used in Virus Total work in a similar manner to NCD (i.e., identify similar patterns). However, a more detailed analysis is needed to determine if these detected files are really false positives or if they are in fact malware. It is also worth noting that the majority of these programs were only flagged as malware by one or two engines out of the 59 engines used in Virus Total. In fact, only 11 programs out of all benign-ware were classified as malware by more than 5 engines. The engine that flagged the most benign-ware as malware for both sets (45 and 47), was only able to detect a small number of malware (35 and 54) in sample set 1 and the diverse set. This engine claims to use a DNA matching algorithm to detect malware which might explain the similarity in false positives to our NCD classifier.

The final two rows of the table show the highest and lowest number of malware and benign-ware that were classified by a single anti-virus engine as malware for each sample set. Again we can see a noticeable difference between the two sample sets in the number of detected malware. The best performing engine in sample set 1 found 883 out of 994 malware (88.8\%), while in the diverse set the best performing engine found 525 out of 980 (53.6\%).  In each set, a different engine had the best performance. Table \ref{tab:topav} shows the false positive, true positive and accuracy rates for these two engines. Engine 1 was the best performing engine in sample set 1 and the second best performing engine in the diverse set. We can see from the results in the table that the false positive rate is around 2\%, however accuracy rates are considerably lower than those obtained by any of our classifiers. 

Engine 2 is the best performing engine in the diverse set but only detected 175 malware in sample set 1. This engine has a very low false positive rate but also suffers from low accuracy. Note that we can achieve low false positive rates in our classifiers by adjusting the voting threshold in the decision forest while achieving higher accuracy rates than those achieved by these two engines.
In both samples, our classifiers achieve considerably higher true positive rates than the combined performance of all 59 anti-virus engines. We also have a lower false positive rate and higher accuracy.   
These results, however, are provided for reference and cannot be used to compare our approach to these engines. The reason is that these results are for the whole set while our approach uses a sub set of programs as feature and training sets. Additionally, as we explained before, we do not have a ground truth and we do not know how these commercial engines work and how much manual effort is involved in building them compared to our approach which is fully automated.

\begin{table}[t]
\centering

   \begin{tabular}{lrrrrrr}
        \toprule                   
	& \multicolumn{3}{c}{Set 1}   & \multicolumn{3}{c}{Diverse Set}  \\ 
           & FP & TP & Acc. & FP & TP & Acc. \\ 
	\midrule
        Engine 1 & 0.019 &0.888  & 0.935 & 0.022 & 0.529 & 0.756  \\
      Engine 2           & 0.006 & 0.176 & 0.586 & 0.005 & 0.536 & 0.768 \\ 
      All Engines &  0.094 & 0.947& 0.926 & 0.105 & 0.805 &0.851 \\
	\bottomrule
       \end{tabular}
   \caption{Scan results of the top performing engines in both sets.}
     \label{tab:topav}
     
  \end{table}

\begin{figure}[t]
 \begin{center}
   \includegraphics[width=0.95\columnwidth]{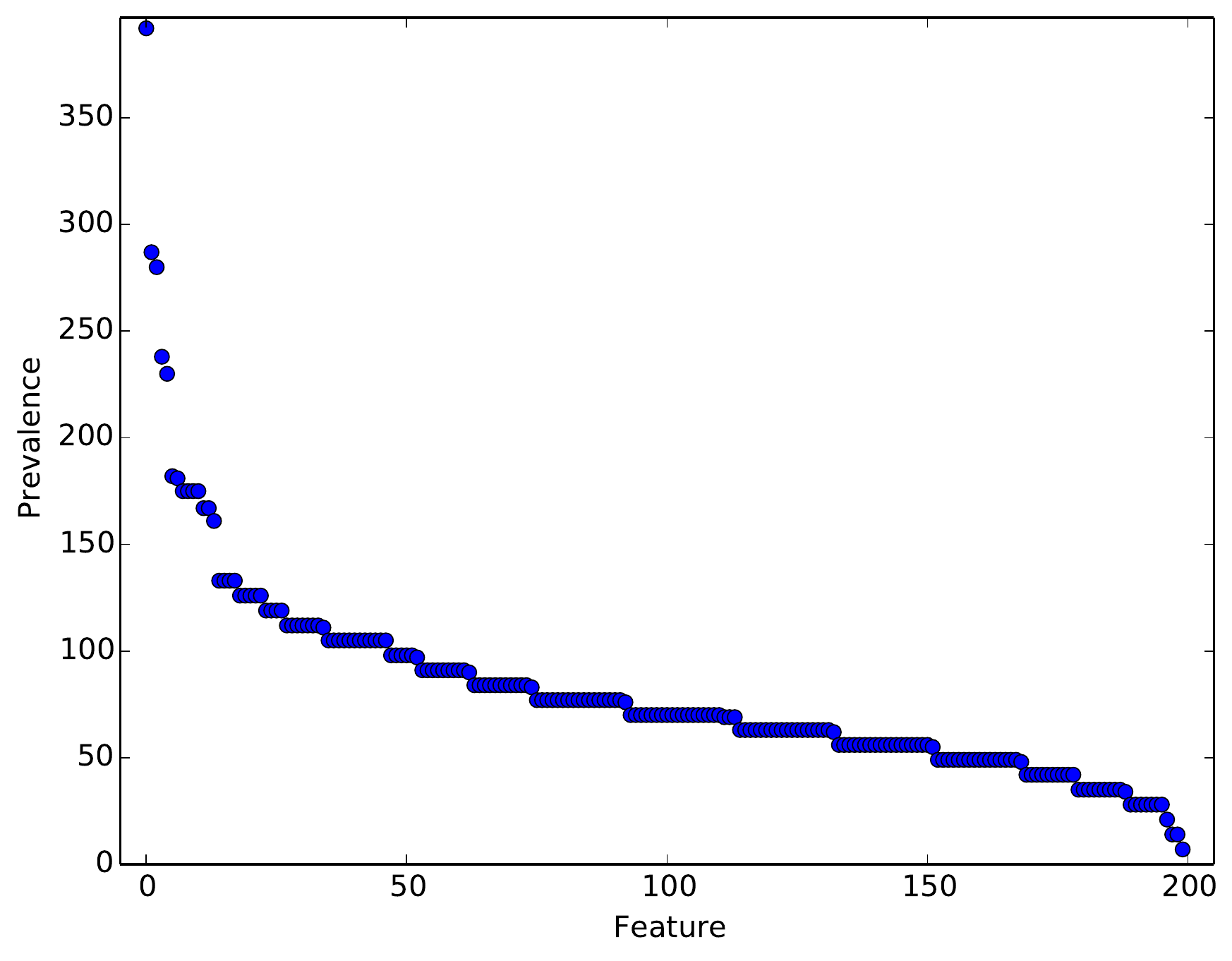}
 \end{center}   
  \caption{Relative feature importance.}
  \label{fig:importance}
\end{figure}

\subsection{Feature Importance Analysis}

We follow the methodology by Breiman~\cite{Breiman2001} to evaluate the importance of specific features, on the basis of how often they are used in the random forest classifier. \autoref{fig:importance} illustrates the importance of each of the 201 features in a single run of the classifier training on sample 1. The features are ordered by importance on the x-axis, with the most used features to the left, and the least used to the right. On the y-axis we plot the prevalence, a proxy for importance, of each feature within the random forest.

It is noteworthy that a bag a of about 10--20 features are relatively important and used quite often. By inspection we note that compressibility is usually within these commonly used features, but is not always the most common one. This illustrates, along with the improved performance, that NCD does add value to classification, and provides valuable features. We also note that the bulk of features provide some value, and we conjecture keeping a varied portfolio makes classification more robust and general.

\section{Related Work}
\label{sec:relwork}

Since NCD has been around since 2003, it is not surprising that there are a number of papers that use or refer to Kolmogorov complexity or information theoretic techniques, not necessarily NCD. These can be grouped into two closely related but different problems: those that consider detection only (as does this paper) and those that consider classification, either in conjunction with the detection problem or by itself..


\subsection{Detection}

Dang, Liu and others use dynamic Markov compression (DMC) to classify binary executables as malware or benign programs. Neither their method nor their results are reported in a clear way \cite{DL+JCIS2011}. They do some light pre-processing by removing headers and white spaces and converting each executable to hexadecimal. For each executable they select substrings of length 1 KB, 10 KB and 100 KB in a manner not explained. This is to reduce the space overhead required in their compression approach. They compress  candidate strings using DMC models for both known benign files and a known set of malware. The candidate is classified according to the smallest compression of the file achieved. They experiment on 2000 benign files and 1000 malware with 15\% of the benign files used for training along with with 30\% of the malware and the remainder in a test set. The results are given as ROC curves but without any discussion, accuracy rates or other information.

Gong, Tan and Zhu devise a detection scenario for malware through the use of compression with Prediction by Partial Matching (PPM) \cite{GTZ+ICISE2009}. They build Markov Models of different malware families but only offer a ``preliminary experiment''. This seems to consist of 200 malware divided into training and test seta in a 30\%--70\% split. However they don't provide any results or descriptions of the experiments.

Abbas and Harris sketch an intrusion detection system that employs compression and hashing \cite{AH+ATNAC2010}. Their system has components that profile worms, profile the network, profile malware, and profile events on the system. All of these profilers employ NCD. The malware profiler is intended to test samples gathered from various sources against various anti-virus vendors to establish some ground truth. then the sample, once identified as malware, is run in a sandbox environment and information is collected. A fuzzy hash is created which acts as a profile. A matrix of NCD values is created for the set of profiles and used to determine a distance threshold which is saved with the database. Collectively, this becomes the malware detector. The report of the ``initial runs'' of the system is not clear with regard to success in malware detection.


\subsection{Classification}

Wehner uses NCD to cluster worms and compressibility to identify packed and encrypted traffic on a network \cite{Weh+JCS2007}. She uses rootless tree diagrams as per the CompLearn tool developed by Cilibrasi \cite{CompLearn} together with discussion of the clustering results. There is no rigorous experimental approach or intention to evaluate NCD as a classifier.

Bailey, Overhead and others use NCD in the context of clustering data derived from five minutes of instrumented execution of malware on a virtual machine \cite{BOA+RAID2007}. Their aim is to address the inconsistencies and variations that arise in labels for malware programs from the various tool specific systems. They argue that a label system for malware files should be consistent and complete and that labels should be concise. They demonstrate that existing systems fail to achieve this by assembling a corpus of 3,700 malware collected between 2004 and 2007 and analysing these files with five different AV tools. They use the Backtracker system to capture event information during the execution. After extracting information of interest, such as files modified, processes created, network connections made, they use this to create an execution profile for each program. By applying NCD pairwise to the profiles they have the raw material for a hierarchical clustering algorithm and they evaluate the results of this with respect to the consistency, completeness and concision of the labels in comparison to those produced by the malware tools. Although their experiments are comparable to ours in scale they are not aiming to detect and classify programs as malware or benign programs.

Apel, Bockermann and Meier experimentally evaluate a set of distance measures for programs with the aim of identifying the most appropriate measure for clustering malware based on behaviours \cite{ABM+MSMB2009}. They consider this as contributing to a detection approach but their experiments only seek to cluster malware rather than classify candidates. They collect 1195 malware samples from honeypot sites and produce traces from these using CWSandbox. They complain about the difficulty of establishing some ground truth with respect to the files. They use the traces to compare and evaluate four distance metrics: Levenshtein distance, approximated edit distance, Normalised Compression Distance, and Manhattan Distance using tries. They rank these distances according to their ability to cluster the traces, measured by the number of continuous (sic) system call sequences that are shared by the traces in the cluster. Their conclusions include that NCD should not be used to analyse malware execution traces.

Gurrutxaga, Arbelaitz et alia also evaluate a set of distance measures as to their suitability for clustering dynamic traces produced by malware \cite{GAM+AusDM2008}. They cite both Bailey et alia and Wehner. Malware is executed for one minute in a controlled environment and information about the execution is collected. They compare different metrics, including NCD, as well as two different ways of representing the collected data. They use three hierarchical clustering algorithms and evaluate the each of these three sets of choices

Wicherski has developed a fast, non-cryptographic hashing algorithm for malware clustering that works on files in Portable Execution format \cite{Wich+LEET2009}. The information hashed includes the file's compressibility ratio as an upper approximation to its Kolmogorov Complexity alongside structural properties such as heap commit size. Again, the aim is to cluster malware rather than develop a detector. He tests his algorithm on two corpora consisting of 184,538 and 90,105 malware and evaluates the resultant clusters by examining the names of the malware and by comparison with clustering on benign programs where he can more easily obtain a ground truth.

Calliat, Desnos and Erra speculate that NCD might be used as a first filtering tactic to select malware from a database of known malware to find those most similar to an unknown malware \cite{CDE+ECIWS2010}. Their paper offers no experimental evidence and it is not clear whether they are aiming to detect or only classify.

Although it is  an application neither of NCD or nor of compression, Baysa, Low and Stamp's investigation of similarity between malware groups produced by metamorphic engines is interesting as a point of comparison as it is entropy based and is applied directly to executables \cite{BLS+JCVHT2013}. Their similarity measure relies on earlier work by Sorokin who developed a comparison technique which compares (executable) files via what he names structural entropy \cite{Sor+JCVHT2011}. This involves using wavelet analysis to split each file into segments of varying entropy levels. Bays et alia use structural entropy comparison as a filtering method then compute a similarity score via the Levenshtein distance between corresponding segments for files that pass the filter. They apply their similarity measure to three groups of malware files, each group produced by applications of a single metamorphic engine, and investigate how well the similarity measure distinguishes the group from a fixed group of 16 benign programs.They report 100\% detection accuracy for the malware groups produced by the G2 and MWOR engines and an accuracy of 0.93539 for the detecting the group produced by the NGVCK engine.

Finally, Symantec's AESOP detection system is an interesting commercial example of leveraging an existing detection and classification tool (in this case Symantec's) via a similarity measure to enable detection on a larger scale while avoiding the need to statically or dynamically analyse every file \cite{TRC+KDD2014}. AESOP relies on the large database owned by Symantec and voluntarily contributed to by users of Norton utilities (Norton Community Watch). Through the simple observation that some machines/users tend to accumulate malware while others do not they can perform a highly light weight analysis of file associations via machines and, starting from the ground truth established by their anti virus tool, classify billions of files within realistic timescales. NCD/compression is a computationally expensive but requires no proprietary database of millions of machines and their files. Furthermore, scalability can be radically improved through the use of a classifier.

\section{Conclusion}
\label{sec:conc}

In future work, we plan on experimenting with selecting those features that
provide the most information, and building cheaper classifiers on this basis.
Reducing the number of features may make classification faster, as NCD only has
to be computed over a smaller number of files, at the possible cost of reducing
robustness and accuracy. Establishing the optimal trade-off between high-value
features and performance is a promising avenue for future work.

As we have shown, the combination of NCD and decision forests is powerful, with
many potential applications.  We have begun applying it to execution traces and
log files.



\bibliographystyle{abbrv}
\bibliography{Bib/triturate}

\end{document}